\documentclass[12pt]{article}

\usepackage{url}
\usepackage{setspace}
\usepackage{amsmath,amssymb,amscd, amsthm}
\usepackage{bbold}
\usepackage{amsfonts}
\usepackage[usenames,dvipsnames]{color}
\usepackage{graphicx}

\usepackage{hyperref}

 \def\p{\partial}
 \newcommand{\bea}{\begin{eqnarray}}
\newcommand{\eea}{\end{eqnarray}}
\newcommand{\be}{\begin{equation}}
\newcommand{\ee}{\end{equation}}
\newcommand{\ba}{\begin{align}}
\newcommand{\ea}{\end{align}}

\newcommand{\ZZ}{\mathbb{Z}} 

\newcommand{\Tr}{{\rm {Tr}}}

\newcommand\rref[1]{(\ref{#1})}

\newcommand{\Fc}{{\cal F}}

\newcommand{\cali}{\includegraphics[height=3.725mm, trim = 0mm 0mm 0mm 0mm, clip]{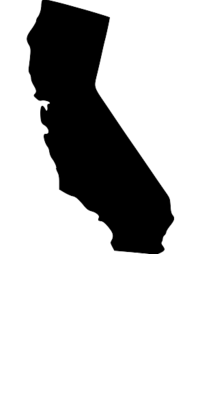}}
\newcommand{\canada}{\includegraphics[height=3.725mm, trim = 0mm 0mm 0mm 0mm, clip]{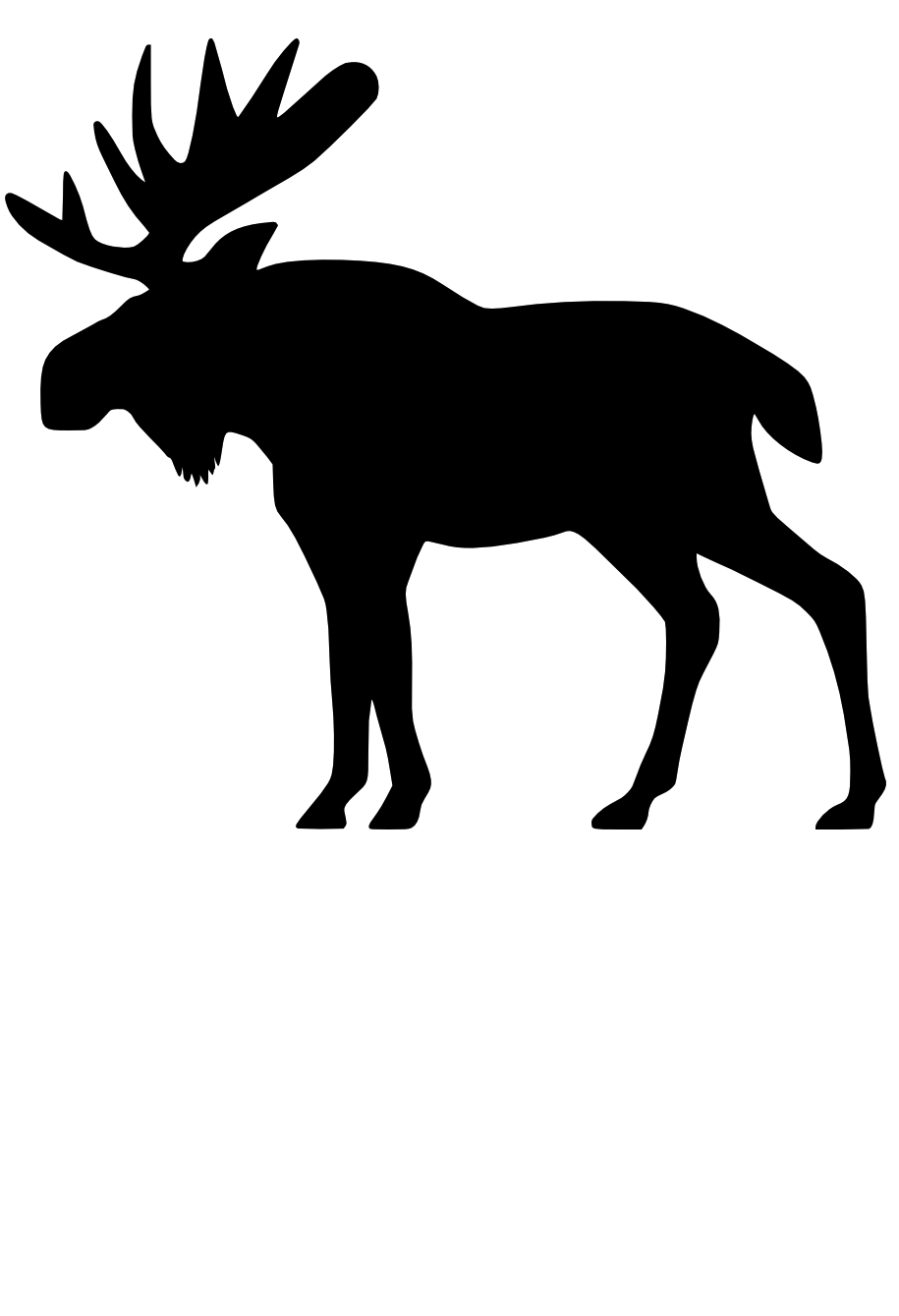}}

\begin{document}

\title{\vspace{-1cm}
	\begin{flushright}\end{flushright}
	\bf{A Cardy Formula for Three-Point Coefficients
		\\ \vspace{2pt}
		{\it or}
		\\ \vspace{6pt}
		How the Black Hole Got its Spots}
	\vspace{12pt}}

\author{
	Per Kraus\cali~
	and
	Alexander Maloney\canada
	\\ \\
	\cali Department of Physics and Astronomy, University of California
	\\
	Los Angeles, CA 90095, USA
	\\
	\\
	\canada Physics Department, McGill University
	\\
	Montr\'eal, QC H3A 2T8, Canada
	\\
	\\
}
\maketitle
%


\vspace{-1em}

\abstract{
Modular covariance of torus one-point functions constrains the three point function coefficients of a two dimensional CFT.  This leads to an asymptotic formula for the average value of light-heavy-heavy three point coefficients, generalizing Cardy's formula for the high energy density of states.  The derivation uses certain asymptotic properties of one-point conformal blocks on the torus.  Our asymptotic formula matches a dual AdS${}_3$ computation of one point functions in a black hole background.   This is evidence that the BTZ black hole geometry emerges upon course-graining over a suitable family of heavy microstates.
}


\clearpage

\tableofcontents

\section{Introduction}

Two dimensional conformal field theories are among the most well-studied quantum field theories.  They describe the dynamics of important statistical and condensed matter systems near criticality, and possess enough symmetry that they can -- in some cases -- be solved exactly
\cite{Belavin:1984vu}.  This has led to the hope that two dimensional CFTs could be completely classified.  This has so far proven impossible, except in the special case of rational CFTs with a finite number of primary operators \cite{Cappelli:1986hf}.  Nevertheless, it is possible to constrain the operator content and dynamics of general irrational CFTs.  Our goal is to describe a new such constraint, and to provide a dual AdS/CFT interpretation involving black hole physics.

The basic dynamical data of a two dimensional CFT is simple to state: every theory is completely determined by the dimensions and three-point function coefficients of the primary operators.  An arbitrary correlation function, as well as the partition function on an arbitrary Riemann surface, can be constructed in terms of this data.
The basic strategy of the conformal bootstrap program is to impose consistency conditions which constrain the allowed dimensions and three-point coefficients.  For 2D CFTs one needs only to impose crossing symmetry of four-point functions on the sphere and modular covariance of one-point functions on the torus; these two conditions are sufficient to imply higher point crossing symmetry and higher genus modular invariance \cite{Moore:1988uz, Sonoda:1988fq}. Recent work has focused primarily on constraints from four-point crossing symmetry, which has led to impressive successes in higher dimensional CFTs; see \cite{Rychkov:2016iqz,Simmons-Duffin:2016gjk} for reviews and references.  Our goal is to intiate a study of the constraints coming from modular covariance of torus one-point functions, which have so far been left out of the fun.

The motivating example for our considerations is Cardy's derivation of the asymptotic density of states of a two dimensional CFT \cite{Cardy:1986ie}.  The starting point is the observation that the partition function of a CFT on the circle $S^1$
\be
Z(\beta) =
\Tr_{{\cal H}_{S^1}} e^{-\beta H} = \sum_i e^{-\beta E_i}
\ee
is invariant under the temperature inversion $\beta \to {4\pi^2 / \beta}$, which is a consequence of modular invariance of the torus partition function.  This $\beta\to 4\pi^2/\beta$ symmetry relates the low temperature behaviour of the theory to the high temperature behaviour.
In particular, it relates the asymptotic density of states at high energy to the energy of the ground state, which -- via the conformal anomaly -- is determined by the central charge.
The energies $E_i$ are, up to a constant shift, equal to the scaling dimensions of the operators on the plane.
So the spectrum of operator dimensions is constrained by modular invariance.

In this paper we will consider instead the finite temperature expectation value of a primary operator $O$:
\be\label{gandalf}
\langle O \rangle_\beta = \Tr_{{\cal H}_{S^1}}O e^{-\beta H} = \sum_i \langle i|O|i\rangle e^{-\beta E_i}~.
\ee
The coefficients $\langle i|O|i\rangle$ in this expansion are essentially equal to the three-point function coefficients $ C_{iiO}$ of $O$ with a complete basis of operators $O_i$.
The one-point function \rref{gandalf} transforms in a known way under the temperature inversion $\beta \to {4\pi^2/\beta}$.  As we will show in section 2, this determines the behaviour of the three-point coefficients $C_{iiO}$ when the dimension $E_i$ is large in terms of the three-point coefficients for low dimension operators.  In particular we will find a universal formula for the {average} value of the three-point function coefficient $C_{iiO}$ as a function of $E_i$. Just as in Cardy's original formula, the asymptotic behaviour of the theory -- in this case the light-heavy-heavy three-point coefficients -- is determined by the dynamics of light operators.

Our result has an interesting dual holographic interpretation.  In the AdS/CFT correspondence every two dimensional conformal field theory can be interpreted as a theory of quantum gravity in three dimensional Anti-de Sitter space \cite{Maldacena:1997re}.  The heavy CFT states are conjectured to be dual to quantum states of black holes in AdS$_3$.  An important piece of evidence is the fact that Cardy's formula for the asymptotic density of states matches precisely the Bekenstein-Hawking formula for the entropy of the corresponding black holes \cite{Strominger:1997eq}.  We are now in a position to take this one step further, and argue that the CFT {observables} -- i.e. the expectation values $\langle i|O|i\rangle$ -- match those of the corresponding black hole.  In section 3 we will compute the one-point function in a black hole background and show that it precisely reproduces our formula for  $\langle i|O|i\rangle$.  Our bulk computation relies on detailed properties of the black hole metric, so demonstrates how certain features of the black hole geometry are visible in CFT observables.  In particular, since the bulk computation matches only the {\it average} value of $C_{Oii}$ we see that the black hole geometry emerges only when we coarse grain over all heavy microstates $|i\rangle$.

In the final section of our paper we will take our asymptotic analysis one step further.
In defining the dynamical data of a conformal field theory one needs only to determine the three-point coefficients of primary operators; descendant operator correlation functions are then fixed using conformal Ward identities.  Our most general formula for three-point coefficients, derived in section 2, gives the asymptotic behaviour of $C_{Oii}$ averaged over all states $|i\rangle$ in the theory, both primaries and descendants.  In section 4 we will understand to what extent we can constrain the asymptotics of primary operator three-point coefficients.  To do so we will need to understand the conformal blocks for torus one-point functions.  In an appendix we will study the asymptotic behaviour of these blocks, extending earlier work \cite{Hadasz:2009db}.  As a result we will derive a similar expression for the asymptotic three-point coefficients of primary operators.  Our formula is valid only when $c\gg 1$, and relies on some other assumptions; we leave the study of ${\cal O}(1/c)$ corrections to future work.

\section{A Cardy formula for three-point coefficients}
\label{sec:Cardy}

Our goal is to derive a Cardy-like formula for the asymptotics of the three-point function coefficients in 2D CFTs.  In this section we will not distinguish between primary and descendant operators; a similar formula for the asymptotics of three-point coefficients of primary operators will be derived, under certain assumptions, in section~\ref{sec:descendants}.

\subsection{Torus one-point functions}

Our central object of interest is the one-point function
of a primary operator $O$ of dimension $(H,\bar H)$ on a torus with modular parameter $\tau$:
\bea\label{onept}
\langle O\rangle_\tau &=& \Tr~ O~ q^{L_0-{c\over 24}} {\bar q}^{{\bar L_0}-{c\over 24}}
\cr
&=& \sum_i \langle i| O |i\rangle q^{\Delta_i-{c\over24}}{\bar q}^{{\bar \Delta_i} - {c\over 24}} 
\eea
where $q=e^{2\pi i \tau}$.  In the second line we have expanded in a basis of states on the cylinder, labelled by an index $i$, which have conformal dimensions $(\Delta_i, {\bar \Delta}_i)$.  We have included explicitly the shift in these dimensions by $c/24$, so that $(\Delta_i, {\bar \Delta}_i)$ are the dimensions of the corresponding operator $O_i$ on the plane.  We will denote by $E_i = \Delta_i + {\bar \Delta}_i$ the total scaling dimension, which is (up to this constant shift) the energy of the state $|i\rangle$ on the cylinder.  Note that although we have not explicitly indicated the dependence on $\bar \tau$, $\langle O\rangle_{\tau}$ is not in general an analytic function of $\tau$.

An important feature of \rref{onept} is that, by translation invariance, $\langle O\rangle_\tau$ is a function only of $\tau$ and not of the location of the operator $O$ on the torus.  Similarly, because the states $|i\rangle$ are  $(L_0, {\bar L}_0)$ eigenstates the coefficient $ \langle i| O |i\rangle$ is a constant.
Indeed, $\langle i| O |i\rangle$ is precisely the three-point function coefficient  for the correlation function $\langle O_i O O_i\rangle $ on the sphere,\footnote{Depending on how one defines the theory on the sphere, a proportionality constant may be present (see section 6.7 of \cite{Polchinski:1998rq}), but this plays no role in our analysis.}
\be
\langle i| O |i\rangle=\langle O_i(\infty,\infty) O(1,1) O_i(0,0)\rangle_{S_2}~,\quad O_i(\infty,\infty) = \lim_{z \rightarrow \infty}  z^{2\Delta_i}\overline{z}^{2\overline{\Delta}_i} O_i(z,\overline{z})~,
\ee
where we are taking a Hermitian basis of operators, $O_i^\dagger = O_i$.

\subsection{Modular invariance}

We will now use the invariance of the conformal field theory on the torus under large conformal transformations, which act on $\tau$ as modular transformations.  The partition function of the theory is invariant under such modular transformations, and the primary operator $O$ transforms with modular weight $(H, \bar H)$.  In particular, under the modular transformation
\be
\tau\to \gamma \tau \equiv {a\tau+b \over c\tau +d}
\ee
the one-point function transforms as a Maass form of weight $(H,{\bar H})$:
\be
\langle O\rangle_{\gamma \tau} =  (c\tau+d)^H (c{\bar \tau} + d)^{\bar H} \langle O\rangle_{\tau}~.
\ee
When $O$ is the identity operator this reduces to the usual modular invariance of the partition function.

We will be interested in the $S$ transformation $\tau \to -{1/ \tau}$, under which
\be\label{smod}
\langle O\rangle_{- 1/\tau} =  (\tau)^H ({\bar \tau})^{\bar H} \langle O\rangle_{\tau} .
\ee
The utility of this formula is that it relates the high temperature behaviour of the theory to the behaviour at low temperature.  For the partition function, this determines the asymptotic density of states of the theory in terms of the dimension of the ground state on the cylinder, which by conformal invariance is set by the central charge.
In the present case, this modular invariance constrains the asymptotics of the three-point function coefficients $\langle i|O|i\rangle$ in the limit where $|i\rangle$ is heavy in terms of the three-point coefficients of light operators.

\subsection{The asymptotic formula}

Let us begin by investigating the behaviour of the one-point function when we take $\tau= i {\beta\over 2\pi}$ with $\beta\to\infty$.  We have
\be\label{gollum}
\langle O\rangle_{i\beta\over 2\pi} = \langle \chi |O |\chi\rangle \exp\left\{- \beta \left(E_\chi-{c\over 12}\right)\right\} +\dots
\ee
where $\chi$ is the the lightest operator with non-vanishing three-point coefficient $ \langle \chi |O |\chi\rangle\ne0$,  and $E_\chi=\Delta_\chi+{\bar \Delta}_\chi$ is the scaling dimension of $\chi$.  The $\dots$ in \rref{gollum} represent terms that are exponentially suppressed as $\beta\to\infty$.\footnote{For simplicity we will assume that this operator is non-degenerate --- if there are multiple operators $\chi_i$ of the same dimension then we must include a sum $\sum_i \langle\chi_i |O|\chi_i\rangle\ne0$. This leaves open the interesting possibility that in some cases the sum might vanish.  This might be the case if, for example, $O$ is a current under which two fields $\chi_i$ carry equal and opposite charges.  In this case one would have to include higher order terms.}  From \rref{smod} we therefore have the high temperature ($\beta\to0$) expansion
\be\label{highT}
\langle O\rangle_{i\beta\over 2\pi} = i^S\langle\chi|O|\chi\rangle \left({2\pi\over \beta}\right)^{E_O} \exp\left\{- {4\pi^2\over \beta} \left(E_\chi-{c\over 12}\right)
\right\}+\dots
\ee
where $E_O=H+{\bar H}$ is the scaling dimension of {$O$} and $S$ is its spin.

We can now compare this result to the expansion \rref{onept}, which we rewrite as
\be \label{frodo}
\langle O\rangle_{i\beta\over 2\pi} = \int dE~ T_O(E)~ \exp\left\{ -\left(E-{c\over 12}\right) \beta \right\}
\ee
where
\be\label{tis}
T_O(E) \equiv \sum_i \langle i|O|i\rangle \delta(E-E_i)
\ee
is the total contribution from operators of dimension $E$.
We note that at high temperatures the integral \rref{frodo} will be dominated by operators with large dimension $E$.
Comparing \rref{frodo} with \rref{highT}, we see that we can write $T_O(E)$ as the inverse Laplace transform
\be
T_O(E) \approx  i^S\langle\chi|O|\chi\rangle \oint~d\beta
\left({2\pi\over \beta}\right)^{E_O}
 \exp\left\{\left(E-{c\over 12}\right) \beta -  \left(E_\chi-{c\over 12}\right){4\pi^2\over \beta} \right\}~.
\ee
At large $E$ this integral is dominated by a saddle point with
\be\label{sadtrump}
\beta \approx 2\pi \sqrt{{c\over 12} - E_\chi \over E-{c\over 12}}+ {E_O\over 2 \left( {E -{c\over 12}}\right)} + \dots
\ee
where $\dots$ are terms which vanish more quickly as $E\to\infty$.  We will restrict our attention to the case where the operator $\chi$ is light -- i.e.  $E_\chi<{c\over 12}$ -- so that the saddle point is real.
The saddle point approximation to the integral gives
\be\label{Tapprox}
 T_O(E) \approx \sqrt{2}\pi
N_O \langle \chi |O|\chi\rangle \left({E - {c\over 12}}\right)^{E_O/2-3/4}
\exp \left\{ 4\pi \sqrt{\left({c\over 12}-E_\chi\right){\left({E - {c\over 12}}\right)}}+\dots\right\}
\ee
where $\dots$ denotes terms which vanish as $E\to\infty$ and the prefactor
\be
N_O =
 i^S\left({c\over 12}- {E_{\chi}} \right)^{-E_O/2+1/4}
\ee
is independent of $E$.     We note that (\ref{Tapprox}) gives a a smeared approximation to (\ref{tis}), where we average over states in an energy window set by the saddle point \rref{sadtrump}.  Since $\beta^{-1} \sim \sqrt{E}$, at high energies we have $\Delta E / E\sim 1/\sqrt{E}$ and the approximation becomes sharp.

Rather than studying the {total} three-point function $T_O(E)$ it is useful to ask what will be the typical value of the three-point function coefficient
$\langle i|O|i\rangle$ for an operator $O_i$ of dimension $E$.
The average value of the three-point coefficient
\be
\overline{\langle E|O|E\rangle} \equiv {T_O(E)\over \rho(E)}
\ee
can be computed using Cardy's formula for the density of states
\cite{Cardy:1986ie}\footnote{We have included here the power law prefactor, for reasons that will become clear below.  As above, this factor comes from the integral over Gaussian fluctuations around the saddle point (as in e.g. \cite{Carlip:2000nv}).}
\be\label{cardy}
\rho(E) \approx \sqrt{2} \pi \left(E - {c\over 12}\right)^{-3/4} \exp\left\{4\pi \sqrt{ {c\over 12} \left(E - {c\over 12}\right)}+\dots  \right\}
\ee
as $E\to \infty$.
The average value of the OPE coefficient of $O$ with two operators of dimension $E$ is
\be\label{lhh}
\overline{\langle E|O|E\rangle}
\approx N_O
\langle \chi|O|\chi\rangle \left(E-{c\over 12} \right)^{E_O/2}
\exp\left\{-{\pi c\over 3}\left(1-\sqrt{1-{12 E_\chi\over c}} \right) \sqrt{ {12 E\over c} - 1}\right\}
\ee
This is the desired asymptotic form for the light-heavy-heavy three-point coefficients. In the next section we will describe the bulk dual interpretation of this formula in terms of AdS$_3$ gravity.

It is worth noting the similarity of \rref{lhh} with the original Cardy formula \rref{cardy}.
In Cardy's original formula, the only data that enters into the leading asymptotic density of states is the central charge (i.e. the dimension of the ground state on the cylinder).  If one wishes to understand subleading contributions to Cardy's formula, however, the result depends on the dimensions of other light operators in the theory.
We have obtained a similar formula, which depends in addition on the data of certain light operators -- in particular, the dimension of the external operator $E_O$ as well as the dimension and three-point coefficient $\langle \chi| O|\chi\rangle$ of the lightest operator to which it couples.

One important difference from Cardy's formula is that the three-point coefficients being studied are not positive definite.  In particular, while the average three-point coefficient vanishes exponentially as $E \to \infty$, individual three-point coefficients might be large. Many of the more precise generalizations of Cardy's formula, such as  \cite{Hellerman:2009bu, Hartman:2014oaa, Friedan:2013cba}, rely on the fact that in Cardy's formula the partition function is a sum of positive definite terms.  It is not clear how to apply these techniques in the present case, where the individual terms in the sum are not positive definite.
In general, it would be interesting to further constrain the statistics of the three-point function coefficients.

So far we have made no assumptions about the value of $c$ in our derivation.  Equation \rref{lhh} could be applied to a Minimal Model CFT, for example.  Simplifications occur if we take $c$ to be large, however, which would be the case when the theory is dual to semi-classical gravity in AdS${}_3$.  In this case an interesting limit is one where the external operator $O$, and the operator $\chi$ to which it couples, are light: $E_O\ll c$, $E_\chi \ll c$ as $c\to\infty$.  In this case
\be\label{lhh2}
\overline{\langle E|O|E\rangle} \approx \tilde{N}_O \langle \chi |O|\chi \rangle
\left({12 E\over c} -1\right)^{E_O/2}
\exp\left\{ - { 2 \pi} E_\chi \sqrt{{12 E\over c}-1 } \right\}~.
\ee

\section{The AdS$_3$ interpretation}
\label{sec:bulk}

We  will now describe the interpretation of the above results in terms of AdS${}_3$ gravity.

A typical finite-$c$ conformal field theory is not expected to be dual to semi-classical bulk gravity.  One might therefore expect that one must  make certain assumptions about the CFT in order to match a bulk derivation, such as large $c$ or a sparseness constraint on the light spectrum or three-point coefficients. This will not turn out to be necessary.  In particular, we will show that, although equations \rref{lhh} and \rref{lhh2} were derived for general CFTs -- assuming only  that $E \gg c$ and the existence of an operator $\chi$ with $\langle \chi|O|\chi\rangle\ne 0$ and $E_\chi< {c\over 12}$ -- these formulas can nevertheless be derived in semi-classical AdS${}_3$ gravity.

A very similar situation arises in Cardy's formula for the density of states, which is derived in a general CFT assuming only that $E\gg c$. Nevertheless it matches the semi-classical Bekenstein-Hawking formula in AdS${}_3$.  We take this as evidence that AdS gravity may capture universal aspects of CFT dynamics beyond the naive regime of validity of semi-classical gravity.

\subsection{The AdS${}_3$ setup}

We begin by giving a schematic bulk derivation of the various terms in our asymptotic formula for three-point coefficients.  The detailed Witten diagram computation  will be deferred to section~\ref{sec:witten}.

We will consider first equation \rref{lhh2}.
We will take $O$ and $\chi$ to be scalar primary operators which are light in the sense that $E_O, E_\chi \ll {c \over 12}$.  So $O$ and $\chi$ are dual to perturbative bulk scalar fields $\phi_O$ and $\phi_\chi$ in AdS${}_3$.   The bulk theory will contain a $\phi_\chi^2 \phi_O$ interaction term with coupling proportional to $\langle \chi|O|\chi\rangle $.

We wish to compute the expectation value of $\langle E| O|E\rangle$ in a typical state with energy $E\gg {c\over 12}$.  At high energy, $|E\rangle$ is well-described by the BTZ black hole geometry \cite{Banados:1992wn}
\be\label{btz}
 ds^2  = - (r^2-r_+^2)dt^2 +{dr^2 \over r^2-r_+^2}+r^2 d\phi^2
 \ee
where $\phi \cong \phi+2\pi$.  We work in units where the AdS radius is ${\ell_{AdS}}=1$.  The area of the horizon $A=2\pi r_+$ is related to the energy by
\be\label{rp}
r_+=\sqrt{{12 E\over c}-1 }~.
\ee
Since $E\gg {c\over 12}$ the area of the black hole is large in AdS units.
Although an individual microstate $|E\rangle$ will not necessarily have a geometric description, the metric \rref{btz} is expected to emerge upon coarse-graining over a suitable family of microstates.  Indeed, in all of our asymptotic formulas we compute only the microcanonical value of the three-point coefficient averaged over all states with with fixed energy.

Let us begin by considering $\langle E| O |E \rangle$ in the limit that the fields $\phi_O$ and $\phi_\chi$ are very massive, so that $E_O\approx m_O$ and $E_\chi\approx m_\chi$ are taken to be much greater than $1$, but still much less than $c$. In this approximation a bulk two-point function for a field of mass $m$ is given by $e^{-m L}$, where $L$ is the geodesic length between two points.
 \begin{figure}[t!]
   \begin{center}
 \includegraphics[width = 0.7\textwidth]{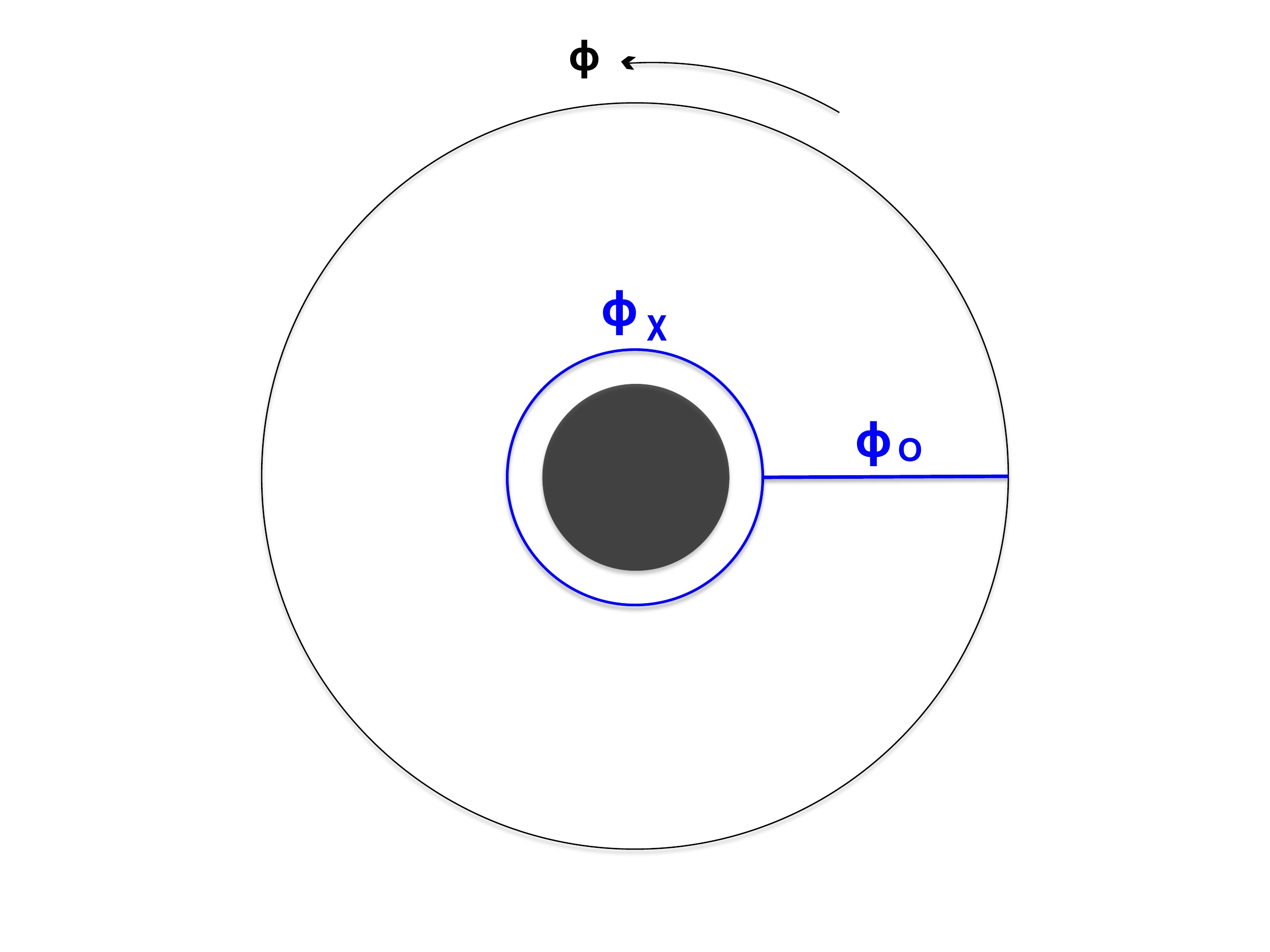}
 \caption{Geodesic approximation to the one-loop contribution to the one-point function of $\phi_O$. A constant time slice of the BTZ metric is depicted.   The $\phi_\chi$ particle wraps the horizon and meets  the $\phi_O$ particle emanating from the boundary at a cubic vertex.  If the geodesic worldlines are replaced by propagators and the cubic vertex is integrated over spacetime, this becomes a full-fledged one-loop Witten diagram.  }
 \label{fig:geods}
 \end{center}
 \end{figure}

We then consider the contribution to $\langle E|O|E\rangle$ sketched in Figure~\ref{fig:geods}: a $\phi_O$ particle propagates from infinity to the horizon, where it splits into a pair of $\phi_\chi$ particles which wrap the horizon.\footnote{A similar process was considered in \cite{Alkalaev:2016ptm}, where it was given a somewhat different interpretation.}
The amplitude for this process is proportional to the cubic coupling $\langle \chi |O|\chi\rangle$.  The $\phi_\chi$ geodesic wrapping the horizon gives a contribution $e^{-m_\chi A}$.  A radial geodesic from the horizon $r_+$ out to a radial coordinate $r=\Lambda$ has length
\bea
L = \int^\Lambda_{r_+} {dr \over \sqrt{r^2-r^2_+}}
&=&
\cosh^{-1}\left({\Lambda \over r_+}\right)
\cr
&\approx& \log \Lambda - \log r_+ + \dots
\eea
where in the second line we have taken $\Lambda\to\infty$.  Discarding the $\log \Lambda$ divergence gives the regularized length which appears in the bulk boundary propagator:
\be
e^{-m_O L_{reg.}} \approx r_+^{m_O}
\ee
where we have neglected terms which are subleading at large $r_+$.
Putting this together we have
\bea\label{bulklhh}
\langle E |O |E \rangle &\approx&
\langle \chi|O|\chi \rangle~ r_+^{m_O} \exp\left\{-2\pi m_\chi r_+\right\}
\eea
Using \rref{rp}, $E_O\approx m_O$ and $E_\chi\approx m_\chi$ this matches precisely our asymptotic formula \rref{lhh2} for the three-point coefficients.

The above derivation assumed that the $O$ and $\chi$ particles were light.  However, in deriving the general asymptotic formula
\rref{lhh} we  assumed only that $E_\chi< {c\over 12}$.
 When $E_\chi$ is of order $c$ the state $|\chi\rangle$ is dual to not to a perturbative field but rather to a massive point particle in AdS${}_3$, which backreacts on the $AdS{}_3$ geometry to give a conical defect geometry \cite{Deser:1983tn,Deser:1983nh}
\be
\label{defect}
ds^2 = -(1+r^2) dt^2 + {dr^2 \over 1+r^2} + r^2 d\phi^2,~~~~~ \phi\cong 
\phi + 2\pi - \Delta \phi
~.
\ee
Here the deficit angle $\Delta \phi$ is related to the mass of the $\chi$ particle by
\be
m_\chi = {c\over 6} {\Delta \phi \over 2\pi}~.
\ee
The important point is that $E_\chi$ should be identified with the ADM mass of \rref{defect} as measured at infinity, which differs from $m_\chi$ due to the gravitational backreaction.  They are related by
\be
m_\chi 
= {c\over 6}  \left(1-\sqrt{1-{12 E_\chi\over c}}\right)
\ee	
Inserting this expression for $m_\chi$ into \rref{bulklhh} reproduces the more general CFT expression \rref{lhh}.

The above arguments are meant to be intuitive  not definitive, and indeed raise some questions.  For example, one might have expected that the geodesic configuration would include an effect due to the radial worldline ``pulling" on the horizon wrapping worldline so as to minimize the total weighted length.  Also, one can ask about how to relax the conditions $E_O, E_\chi \gg 1$.   These issues are addressed by the more careful Witten diagram derivation of \rref{lhh2} given in the next section.  On the other hand,   we  leave a more systematic derivation of \rref{lhh}, which will involve the dynamics of conical singularities in the BTZ background, to future work.

Before proceeding let us make a few comments on the interpretation of our derivation \rref{bulklhh}.  First, we note that the BTZ metric \rref{btz} should be regarded as an effective description of the state $|E\rangle$ which emerges only when we coarse grain over many states at fixed energy.  An individual microstate $|E\rangle$ may contain large fluctuations which deviate significantly from \rref{btz}, and may not even have a metric description.  Our asymptotic formula \rref{lhh2} computes only the average value of $ \langle E |O |E \rangle $ in the microcanonical ensemble, where we average over all states at fixed energy.  Indeed, the states over which we average are counted by Cardy's formula \rref{cardy}, which is the Bekenstein-Hawking formula that counts black hole microstates \cite{Strominger:1997eq}.  Our bulk formula \rref{bulklhh} relied on detailed properties of geodesics in the black hole geometry.
We can therefore interpret our asymptotic formula \rref{lhh2} as further CFT evidence that the black hole geometry emerges upon course graining over microstates.

Another point worth mentioning is that the nonzero one-point function of $O$ is a consequence of Hawking radiation in the bulk.  This feature can be made manifest by writing down the thermal state for $\phi_\chi$, using this to compute $\langle \phi_\chi^2\rangle$, and then thinking of this as a source for $\phi_O$.  This approach leads to the same result at the end.

\subsection{Witten diagram calculation}
\label{sec:witten}

 To do the computation properly we must compute the appropriate Witten diagram in
 the BTZ metric \cite{Banados:1992wn} 
 \be
 \label{ebtz}
 ds^2  = (r^2-r_+^2)dt_E^2 +{dr^2 \over r^2-r_+^2}+r^2 d\phi^2
 \ee
 which we write in Euclidean signature, with $t_E\cong t_E +2\pi/r_+$.
 The leading contribution to the one-point function is the 1-loop diagram with a
 single cubic vertex depicted in Figure~\ref{fig:geods}, but now with the worldlines replaced by propagators and the cubic vertex integrated over the BTZ spacetime:
 %
 \be\label{witten}
 \overline{\langle E|O|E\rangle }= C_{O\chi\chi} \int\! dr dt_E d\phi~r~
 G_{bb}(r;\Delta_\chi) G_{b\p}(r,t_E,\phi;\Delta_O)
 \ee
 where we have used translation invariance to place the boundary operator at
 $t_E=\phi=0$.  Here $G_{bb}$ and $G_{b\partial}$ are the bulk-bulk and
 bulk-boundary propagators in the the BTZ geometry \rref{ebtz}.
 We will consider the case of scalar operators with scaling dimensions $\Delta_O$ and $\Delta_\chi$, which are equal to the energies $E_O$ and $E_\chi$ of the corresponding states on the cylinder.

We are interested in the large $r_+$ limit, where the integral is tractable.
 We will absorb all $r_+$ independent factors into the normalization $C_{O\chi\chi}$
 of the three-point vertex
 and show that \rref{witten} is
 \be
 \overline{\langle E|O|E\rangle }\approx  C_{O\chi\chi}~
 r_+^{\Delta_O}~\exp\left\{-2\pi \Delta_\chi r_+\right\}~,\quad r_+ \rightarrow
 \infty
 \ee
 This reproduces our CFT result \rref{lhh2} for the asymptotic three-point coefficients.

 \subsubsection*{The bulk-bulk propagator}

 The AdS${}_3$ propagator for a scalar field of mass $m^2=\Delta(\Delta-2)$ is \cite{D'Hoker:2002aw}
 \be
 \label{adsprop}
 G^{AdS}_{bb}(y,y')=-{1\over 2\pi} {e^{-\Delta \sigma(y,y')}\over
 	1-e^{-2\sigma(y,y')}}
 \ee
 where $\sigma(y,y')$ is the geodesic distance.
 The propagator obeys $(\nabla^2 -m^2 )G_{bb}(y,y')={1\over
 \sqrt{g}}\delta^{(3)}(y-y')$.
The BTZ geometry is a quotient AdS${}_3/\ZZ$, so we can use the method of images to
obtain the BTZ propagator \cite{KeskiVakkuri:1998nw}
\be
\label{btzprop}
G_{bb}(y,y')=-{1\over 2\pi} \sum_{n\in \ZZ} {e^{-\Delta \sigma_n(y,y')}\over
	1-e^{-2\sigma_n(y,y')}}
\ee
where $\sigma_n(y,y')$ is the geodesic distance between $y$ and the $n^{th}$ image
of $y'$ under the BTZ identification.  In terms of the metric \rref{ebtz}, the
$\ZZ$ identification is the periodic identification $\phi\cong\phi+2\pi$.  The sum
over $n$ in \rref{btzprop} can be interpreted as a sum over geodesics in the BTZ
background: $\sigma_n(y,y')$ is the length of a geodesic which wraps $n$ times
around the event horizon.

We are interested in $G_{bb}(y,y)$, where the two bulk points are at the same
location.  The $n=0$ term in \rref{btzprop} gives a naively divergent contribution
to the one-point function \rref{witten}; this is the usual UV divergent tadpole
contribution, present even in empty AdS${}_3$, which is cancelled by a local
counterterm.  The $n\ne 0$ terms give non-trivial, finite contributions to
\rref{witten}.  This subtracted two point function depends only on the radial
location $r$ of the point $y$, and is
\be
G_{bb}(r) = -{1\over \pi} \sum_{n=1}^\infty {e^{-\Delta \sigma_n(r)}\over
	1-e^{-2\sigma_n(r)}}~.
\ee
Here $\sigma_n(r)$ is the length of the closed geodesic which starts at radial
location $r$ and winds $n$ times around the horizon before returning back to the
starting point.

We now note that, since the horizon itself is a geodesic, $\sigma_n(r_+)=2\pi n
r_+$.  And clearly $2\pi n r_+ \leq \sigma_n (r) \leq 2\pi n r$.  Thus we can
focus only on the $n=1$ term  
\be G_{bb}(r) \approx -{1\over \pi} e^{-\Delta \sigma_1(r)} \ee We now determine the behavior of  $\sigma_1(r)$ as $r_+\to \infty$.

The geodesic corresponding to $\sigma_1(r)$ lies at constant $t$ and so can be parametrized as $r(\phi)$.  Constant $t$ geodesics are governed by the equation %
\be
\left(dr\over d\phi\right)^2 -{r^2(r^2-r_+^2)(r^2-r_0^2)\over r_0^2}=0
\ee %
where $r_0$ denotes the turning point, which can be chosen freely.  We note that there is only a single turning point for $r>r_+$, hence geodesics with both endpoints at radial location $r$ are described by curves that bounce off the turning point at $r_0$. As $r_+\rightarrow \infty$ it's easy to see that we need to take $r_0 \approx r_+$ to satisfy the boundary conditions.  Further, in this regime the geodesic spends almost all of its ``time" near $r=r_+$, apart from fixed length segments where it travels out to the endpoints.   Hence we conclude that $\sigma_1(r) \approx 2\pi  r_+$ and so
\be G_{bb}(r)\approx -{1\over \pi} e^{-2\pi \Delta r_+}~.
\ee

\subsubsection*{The bulk-boundary propagator}

The bulk-boundary propagator for a scalar of mass $m^2 =\Delta(\Delta-2)$ in BTZ can be obtained from the bulk-bulk propagator by taking one of the bulk points to the boundary, and is given by
\be\label{bulkbnd} G_{b\partial}(r,t_E,\phi) = N \left( -\sqrt{{r^2\over r_+^2}-1}\cos(r_+t_E)+{r\over r_+}\cosh(r_+\phi)\right)^{-\Delta} \ee %
where the normalization constant is $N= {1\over 2^\Delta \pi}(\Delta-1)r_+^\Delta$. $N$ is chosen such that $G_{b\partial}(r,t,\phi) \approx r^{\Delta-1}\delta(t)\delta(\phi)$ as $r\rightarrow \infty$.   The boundary point has been taken as $t=\phi=0$.  Properly, in (\ref{bulkbnd}) we should include a sum over images, $\phi \rightarrow \phi+2\pi n$, but we omit this since in our computation the $n\neq 0$ terms are exponentially small as $r_+\rightarrow \infty$.

\subsubsection*{Witten diagram}

With these results in hand, the 1-loop Witten diagram becomes
\be
\overline{\langle E|O|E\rangle }\approx -{\Delta_O-1 \over 2^{\Delta_O} \pi^2}
r_+^{\Delta_O}e^{-2\pi \Delta_\chi r_+} \int\! dr dt_E d\phi ~r
\left(-\sqrt{{r^2\over r_+^2}-1} \cos (r_+t_E) +{r\over r_+}\cosh(r_+
\phi)\right)^{-\Delta_O}
\ee
Now rescale the coordinates,
\be
\overline{\langle E|O|E\rangle }\approx -{\Delta_O-1 \over 2^{\Delta_O}
	\pi^2}r_+^{\Delta_O} e^{-2\pi \Delta_\chi r_+} \int\! d\hat{r} d\hat{t}_E
d\hat{\phi} ~\hat{r} \left(-\sqrt{{\hat{r}^2}-1} \cos (\hat{t}_E) +{\hat{r}}\cosh(
\hat{\phi})\right)^{-\Delta_O}
\ee
so that $\hat{t}_E\cong \hat{t}_E+2\pi$ and $\hat{\phi}\cong \hat{\phi}+2\pi r_+$.  In this form, the only dependence on $r_+$ in the integral comes from the $\hat{\phi}$ integration range. But the integrand is exponentially small for $|\hat{\phi}| \sim r_+$ and so we can freely extend the $\hat{\phi}$ integral over the real line.  Hence the leading term in the integral is $r_+$ independent, and we can write %
\be \overline{\langle E|O|E\rangle } \approx C_{O\chi\chi} r_+^{\Delta_O}e^{-2\pi \Delta_\chi r_+} \ee %
for some $C_{O\chi\chi}$ which is independent of $r_+$.  As advertised, this reproduces \rref{lhh2}.

\ 

\section{Three-point function coefficients for primary operators}
\label{sec:descendants}

In section~\ref{sec:Cardy} we studied the asymptotics of the light-heavy-heavy three-point function coefficient for arbitrary operators.  We now consider the three-point coefficients of primary operators, from which descendant operator three-point coefficients can be derived using Virasoro Ward identities.  The primary operator three-point function coefficients can be regarded -- along with the central charge and the primary operator dimensions -- as the independent data which define a conformal field theory.  We will see that certain assumptions are necessary in order to obtain similar asymptotic formulae.

\subsection{Torus blocks}

We consider, as in section~\ref{sec:Cardy}, the one-point function of a primary operator $O$ on a torus with modular parameter $\tau$:
\bea
\label{blocky}
\langle O\rangle_\tau
&=& \sum_i q^{\Delta_i-{c\over24}}{\bar q}^{{\bar \Delta_i} - {c\over 24}} \langle i| O |i\rangle \cr
&=&
\sum_\alpha \langle\alpha| O|\alpha\rangle q^{\Delta_\alpha-{c\over 24}} {\bar q}^{{\bar \Delta_\alpha}-{c\over 24}} { F}^H_{\Delta_\alpha,c}(q) {  F}^{\bar H}_{\bar \Delta_i,c} ({\bar q})~.
\eea
In the second line we have
written the one-point function as a sum over primaries $|\alpha\rangle$
of dimension $(\Delta_\alpha, {\bar \Delta}_\alpha)$.
The functions ${ F}^H_\Delta(q)$ and ${ \bar F}^{\bar H}_{\bar \Delta} ({\bar q})$
are known as torus one-point function conformal blocks, and encode the contributions of the Virasoro descendants of the primary $|\alpha\rangle$.  These conformal blocks depend only on the dimensions $(H,\bar H)$ and $(\Delta, {\bar \Delta})$ and the central charge $c$.  We will restrict our attention to the case $c>1$.

The torus block ${ F}^H_{\Delta,c}(q)$ can be computed algorithmically using the Virasoro algebra (see e.g. \cite{Hadasz:2009db} for a nice review).  For example, one can compute the coefficients in the $q$ expansion explicitly:
\be
{ F}^H_{\Delta,c}(q) = 1 + \left({H(H-1)\over 2 \Delta }+1\right) q + {\cal O}(q^2)
\ee
Exact closed form expressions for ${ F}^H_\Delta(q)$ exist only in special cases.  One example is $H=0$, when the external operator is the identity.  In this case the one-point function is just the partition function, to which all of the descendant states contribute equally.  This implies that the conformal block is just the Verma module character\footnote{In fact, as discussed in Appendix~\ref{sec:blocks} this expression holds when $H=1$ as well.}
\be\label{vacblock}
{ F}^0_\Delta(q) = \prod_{n=1}^\infty (1-q^n)^{-1} = {q^{1/24} \over \eta(q)}
\ee
which counts the states in the Verma module built on the primary $|\alpha\rangle$.

The coefficients in the $q$-expansion of a conformal block are  polynomials in the external dimension $H$ and rational functions of $\Delta$ and $c$, whose poles and residues are determined by the null vectors of the Virasoro algebra.
This leads to recursion relations that can be used to efficiently compute the blocks explicitly, just as with four-point blocks on the sphere \cite{Zamo:lectures,zamo}.
These recursion relations also allow one to study the blocks in various limits. For example, as noted in \cite{Hadasz:2009db}, we have
\be\label{largeDblock}
{ F}^H_{\Delta,c}(q) = {q^{1/24} \over \eta(q)} + {\cal O}(\Delta^{-1})
\ee
as $\Delta\to \infty$.  The derivation of this formula and its subleading corrections are discussed in detail in Appendix~\ref{sec:blocks}.
Equation \rref{largeDblock} can be understood as the statement that, when the internal operator {$|\alpha\rangle$} is heavy one can regard the external operator ${ O}$ as a small perturbation to the result \rref{vacblock}.
In appendix~\ref{sec:blocks} we will study the regime of validity of this approximation and show that \rref{largeDblock} holds when
\be\label{validity}
\Delta |\log q|^2 \gg 1~.
\ee


\subsection{Asymptotics of three-point coefficients}

We can now derive an expression for the primary operator three-point coefficients.
We will begin by rewriting equation \rref{blocky} as
 \be \label{bilbo}
 \langle O\rangle_{i\beta\over 2\pi} = \int d\Delta d{\bar \Delta}~ T^p_O(\Delta, {\bar \Delta})~ \exp\left\{ -\left(E-{c\over 12}\right) \beta \right\}
  F^H_\Delta(e^{-\beta})F^H_{\bar \Delta}(e^{-\beta})
 \ee
 where
 \be\label{tpis}
 T^p_O(\Delta, {\bar \Delta}) \equiv \sum_\alpha { \langle \alpha|O|\alpha\rangle} \delta(\Delta -\Delta_\alpha)\delta({\bar \Delta} -{\bar \Delta}_\alpha)
 \ee
 is the contribution from primaries of dimension $(\Delta, {\bar \Delta}) $, and  $E=\Delta+{\bar \Delta}$.
We again wish to study the high temperature limit, using \rref{gollum}:
\be
\langle O\rangle_{i\beta\over 2\pi} = \langle \chi |O |\chi\rangle \exp\left\{- \beta \left(E_\chi-{c\over 12}\right)\right\} +\dots
\ee
to constrain the asymptotics of three-point function coefficients.  As in section~\ref{sec:Cardy} the operator $\chi$ is the lightest one with $ \langle \chi |O |\chi\rangle\ne0$ and we assume that $E_\chi<{c\over 12}$.

We must now ask under what circumstances the integral \rref{bilbo} will be dominated by terms with large $\Delta$ and ${\bar \Delta}$.  In section~\ref{sec:Cardy} this was an automatic consequence of the divergence of \rref{gollum} at $\beta\to0$.
Once the conformal blocks are included, however, this is not necessarily the case.  For example, when $c<1$ there are only a finite number of primary states and the integral \rref{bilbo} has a finite range. In this case the divergence \rref{gollum} will come from the behaviour of the conformal blocks $F^H_\Delta(e^{-\beta})$ as $\beta\to0$.  While this is an interesting special case, our primary interest is in matching with AdS${}_3$ gravity at large $c$.

In a generic large $c$ conformal field theory we expect that the small $\beta$ behavior of \rref{bilbo} is controlled by a saddle point at large $\Delta$ and ${\bar \Delta}$.  The existence of a saddle point in the sum over all states is essentially equivalent to the existence of a thermodynamic limit in which macroscopic observables like the total energy assume sharply defined values, with the energy growing with the temperature.      In general, this does not imply that the dimensions of the contributing primaries behave similarly, since the contribution of descendant states must be taken into account; indeed  for $c<1$ theories with a finite spectrum of primaries it must be that the descendants dominate.   However, at large $c$ the asymptotic density of primary states grows rapidly at high energy -- in fact, as we will see below it is essentially given by Cardy's formula.    So in a generic large  $c$ theory we expect the small $\beta$ asymptotics to be controlled by a saddle point at large $(\Delta, {\bar \Delta})$.

Let us therefore proceed by investigating the large $(\Delta, {\bar \Delta})$ asymptotics of \rref{bilbo},  assuming that $c\gg 1$ so that the integral has support at large $E$.
Using \rref{largeDblock} we have
 \bea
  F^H_\Delta(e^{-\beta})F^H_{\bar \Delta}(e^{-\beta})
 &=& \left|{q^{{1/ 24}}\over \eta(\tau)}\right|^2+\dots
\eea
where $\dots$ denote terms which are suppressed as $(\Delta, {\bar \Delta})\to\infty$.
We can then use the $\beta\to 0$ asymptotics of the eta function to obtain
\bea\label{Fapprox}
\textbf{}
F^H_\Delta(e^{-\beta})F^H_{\bar \Delta}(e^{-\beta})
 \approx
 \beta \exp\left\{{1\over 12} \left(\beta + {4\pi^2 \over \beta} \right)\right\} +\dots
 \eea
We note from \rref{validity} that this asymptotic form holds only if we take $(\Delta, {\bar \Delta})\to\infty$ and $\beta\to0$ with  $\Delta  \beta^2\gg 1,~ {\bar \Delta} \beta^2\gg 1$.  We will justify this condition below.

We now sum over spins by writing $T^p_O(E) = \int\! ds T^p_O(\Delta,{\bar \Delta})$, where $E=\Delta+{\bar \Delta}$ and $s=\Delta-{\bar \Delta}$.
We can then  proceed to compute $T^p_O(E)$ as in section~\ref{sec:Cardy}.
We find a nearly identical expression for the inverse Laplace transform: 

 \be
T^p_O(E) = {i^S\over 2\pi} \langle\chi|O|\chi\rangle \oint~d\beta
\left({2\pi\over \beta}\right)^{E_O+1}
 \exp\left\{\left(E-{c-1\over 12}\right) \beta -  \left(E_\chi-{c-1\over 12}\right){4\pi^2\over \beta} \right\}
\ee
This integral has a saddle point with
\be\label{saddle}
\beta \approx 2\pi \sqrt{{c-1\over 12} - E_\chi \over E-{c-1\over 12}}+ {E_O+1\over 2 \left( {E -{c-1\over 12}}\right)} + \dots
\ee
giving the saddle point result
\be
\label{tpis}
T^p_O(E) \approx
\sqrt{2}\pi N^p_O \langle \chi|O|\chi\rangle \left({E - {c-1\over 12}}\right)^{E_O/2-1/4}
\exp \left\{ 4\pi \sqrt{\left({c-1\over 12}-E_\chi\right){\left({E - {c-1\over 12}}\right)}}+\dots\right\}
\ee
with
\be
N^p_O =
 {i^S\over 2\pi} \left({c-1\over 12}- {E_{\chi}} \right)^{-E_O/2-1/4}~.
\ee

We can now ask whether our approximation \rref{Fapprox} was justified.  From \rref{saddle} we see that as long as we take
\be
E \beta^2 \approx {c-1\over 12} - E_\chi \gg 1
\ee
our approximation is valid.
We conclude that our result \rref{tpis}
holds whenever we take $c$ large with $E_\chi\ll {c\over 12}$.  In particular, this result holds when $\chi$ is dual to either a perturbative bulk field or a conical deficit in AdS${}_3$.

We can now compute the average value of the three-point coefficient
\be
\overline {\langle E|O|E\rangle}_p \equiv {T^p_O(E)\over \rho^p(E)}
\ee
where
\be\label{cardy}
\rho^p(E) \approx \sqrt{2} \pi \left(E - {c-1\over 12}\right)^{-1/4} \exp\left\{4\pi \sqrt{ {c-1\over 12} \left(E - {c-1\over 12}\right)}+\dots  \right\}
\ee
is the density of state of primary operators of dimension $E$ in a theory with $c>1$ as $E\to\infty$.\footnote{This version of Cardy's formula can be derived by, for example, taking $O$ to be the identity operator in \rref{tpis}.}   At large $c$ this formula is -- at leading order in $c$ -- indistinguishable from the original Cardy formula which counts all states, not just primaries. 
The average value of the OPE coefficient for $O$ with two primary operators of dimension $E$ is then
\be\label{lhhb}
\overline {\langle E|O|E\rangle}_p
\approx
N^p_O\langle \chi|O|\chi\rangle \left(E-{{\hat c}\over 12}\right)^{E_O/2}
\exp\left\{-{\pi\hat{c}\over 3}\left(1-\sqrt{1-{12 E_\chi\over {\hat c}}} \right) \sqrt{ \left({12E\over \hat{c}} - 1\right)}\right\}
\ee
where ${\hat c}=c-1$.

Our expressions for the asymptotics of the primary operator three-point coefficients are nearly identical to those for the general three-point coefficients -- the only difference is a shift of $c\to {\hat c}=c-1$.  This shift is invisible in the semi-classical large $c$ limit, so the bulk derivations in section~\ref{sec:bulk} continue to apply in this case.  Indeed, the fact that descendant states lead to an effective shift of the central charge has been observed in other contexts (see e.g. \cite{Maloney:2007ud,Keller:2014xba,Benjamin:2016aww}). This shift can be viewed as a one-loop renormalization of the bulk effective central charge due to the presence of Virasoro descendants, whose density of states grows like that of a CFT with central charge $1$.

\section*{Acknowledgements}

We are grateful to John Cardy, Tom Hartman, Henry Maxfield, Gim Seng Ng and Stephen Shenker for useful conversations.  P.K. is supported in part by NSF grant PHY-1313986.
A.M. is supported by the National Science and Engineering Council of Canada and by the Simons Foundation.

 \appendix

 \section{Asymptotics of torus conformal blocks}
 \label{sec:blocks}

 In this appendix we will discuss the asymptotics of one-point conformal blocks on the torus with the goal of understanding the regime of validity of the asymptotic formula \rref{largeDblock}.

 \subsection{Setup}

We will follow the notation of \cite{Hadasz:2009db}.  We consider an external primary operator $\phi_{\lambda,{\overline \lambda}}$ of dimension $(\Delta_\lambda, {\overline \Delta}_\lambda)$ and expand the torus one-point function as a sum over primaries
\be
\label{es}
	\langle \phi_\lambda \rangle =
 	 \sum_{\Delta,\overline{\Delta}}
 	 	C^{\lambda,\overline{\lambda}}_{\Delta,\overline{\Delta}}
 	 	\Fc_\Delta(q)\Fc_{\overline{\Delta}}({\bar q})
 	\ee
 The conformal block $\Fc_\Delta(q)$ depends on $\Delta_\lambda$ and $c$ as well as $\Delta$, but we will suppress the dependence on the former for notational simplicity.
 Here we label primary operators $\nu_{\Delta; \overline{\Delta}}$ by their dimensions $(\Delta, \overline{\Delta})$ and
 \be
 C^{\lambda,\lambda}_{\Delta,\overline{\Delta}}
 =
 \langle \nu_{\Delta; \overline{\Delta}}|\phi_{\lambda,{\overline \lambda}}(1)|\nu_{\Delta; \overline{\Delta}}\rangle
 \ee
 is the OPE coefficient of $\phi_{\lambda,{\bar \lambda}}$ with $\nu_{\Delta, \overline{\Delta}}$.

 A general descendant state $|\nu_{\Delta,N; \overline{\Delta},\overline{N}}\rangle$ will be labelled by a pair $N=\{n_i\}, {\overline N} = \{{\overline n}_i\}$ of sequences of non-negative integers, as
 \be
|\nu_{\Delta,N; \overline{\Delta},\overline{N}}\rangle
=\prod_{i=1}^\infty \left(L_{-i}\right)^{n_i} \left({\overline L}_{-i}\right)^{{\overline n}_i}|\nu_{\Delta; \overline{\Delta}}\rangle
 \ee
The OPE coefficient of $\phi_{\lambda, {\overline \lambda}}$ with a descendant state will take the form
 \be
 \langle \nu_{\Delta,M; \overline{\Delta},\overline{M}}|\phi_{\lambda,\lambda}(1)|\nu_{\Delta,N; \overline{\Delta},\overline{N}}\rangle =\rho(\nu_{\Delta,N},\nu_\lambda,\nu_{\Delta,M})
 	\rho(\nu_{\overline{\Delta},{\bar N}},\nu_{\lambda},\nu_{\overline{\Delta},{\bar M}}) C^{\lambda,\lambda}_{\Delta,\overline{\Delta}}
 \ee 
  To compute the conformal block we must normalize the descendant states approproiately, so will need the Gram matrix
 \be
 [B^n_\Delta]_{MN} \equiv\langle \nu_{\Delta,N}|\nu_{\Delta,M}\rangle
 \ee
The entries in the Gram matrix are rational functions of $\Delta$ and $c$.
We will can then expand the conformal block as
\be
\Fc_\Delta(q)=
q^{\Delta-{c\over 24}} \sum_{n=0}^\infty q^n F^n_\Delta
\ee
where the coefficient
\be
		F^n_\Delta =
		\sum_{|M|=|N|=n}\rho(\nu_{\Delta,N},\nu_\lambda,\nu_{\Delta,M})[B^n_\Delta]^{MN}
	\ee
gives the contribution from descendant states of level $n$.
Here $[B^n_\Delta]^{MN} $ is the inverse Gram matrix at level $n$.
We note that the $\rho(\nu_{\Delta,N},\nu_\lambda,\nu_{\Delta,M})$ are polynomials in $\Delta_\lambda$ which are determined by the Virasoro Ward identities, while $[B^n_\Delta]^{MN} $ is a rational function of $\Delta$ and $c$.

\subsection{Large  $\Delta$ limit}

We now study the large $\Delta$ limit of the coefficients $F^n_\Delta$.

Let us consider the computation of the coefficient
$\rho(\nu_{\Delta, N}, \nu_{\lambda},\nu_{\Delta, M})$ at level $n$.
To compute this we must move the various raising operators $L_{-i}$ appearing in
$| \nu_{\Delta, M}\rangle =  \prod L_{-i}^{n_i}|\nu_\Delta \rangle$ to the left through $\phi$ in order to act on $\langle \nu_{\Delta, M}|$.
Powers of $\Delta$ come from factors of $L_0$ acting on the external states.  These $L_0$ factors come from commutators $[L_m, L_{-m}] = 2m L_0$.
For example, suppose we wish to compute the matrix element $\langle \Delta | L_m \phi_\lambda L_{-m}|\Delta\rangle$ using
\be
\phi_\lambda L_{-m} = L_{-m} \phi_\lambda +[\phi_\lambda,L_{-m}]
\ee
The first term has an $L_{-m}$ which can then appear in $[L_m,L_{-m}]$ to give a factor of $\Delta$.  The second commutator term can be expressed in terms of $\phi_\lambda$ and its derivatives using \rref{Lphi}, leaving no Virasoro generators.  To maximize powers of $\Delta$ we therefore want the fewest number of such commutators.  Thus the leading contribution at large $\Delta$ occurs when we ignore the commutator
$[\phi_\lambda,L_{-m}]$.
  This argument works for a general descendant as well: at leading order in $\Delta$, we can simply move all of the $L_{-i}$ in $| \nu_{\Delta, M}\rangle $ to the left ignoring the commutator with $\phi$.  The result is that the matrix element
\be
\rho(\nu_{\Delta, N}, \nu_{\lambda},\nu_{\Delta, M})= [B^n_\Delta]_{MN}
\ee
is just the usual Gram matrix.
This means that at large $\Delta$ the coefficient $F_{\Delta}^n$ just counts the number of descendant states at level $n$:
\be
F_\Delta^n = p(n) + {\cal O}(\Delta^{-1})
\ee
where $p(n)$ is the number of partitions of $n$.
This leads to equation \rref{largeDblock} for the conformal block
\be\label{largeDFn}
{ F}_\Delta =  \prod_n (1-q^n)^{-1} +  {\cal O}(\Delta^{-1})
\ee

In fact, one can show that equation \rref{largeDFn} is {\it exact} when the external operator has dimension $\Delta_\lambda=0$ or $\Delta_\lambda=1$:
\be \label{shockingnews}
F_\Delta(q) = \prod_n (1-q^n)^{-1} = {q^{-1/24}\over \eta(q)}~,\quad{\rm when}\quad  \Delta_\lambda = 0, 1
\ee
The $\Delta_\lambda=0$ case is clear, since in this case $\phi_\lambda$ is the unit operator. Matrix elements of the identitiy just count states, and there are $p(n)$ states at level $n$.  The $\Delta_\lambda=1$ can be similarly understood by noting that, by translation invariance, we can integrate the operator $\phi_\lambda(z)$ over the unit circle and divide by $2\pi$ to obtain the same result as $\phi_\lambda(1)$.  However, since $\Delta_\lambda=1$ the result is conformally invariant, so commutes with all the $L_n$.  Thus the three-point function of $\phi_\lambda(1)$ in a descendant state is the same as in the original primary.  Indeed, if we consider the case where $\phi_\lambda$ is a conserved current the resulting integral is just the conserved charge.  The statement \rref{shockingnews} is just the statement that every state in the same conformal family has the same charge.

We wish to go one step farther and understand the subleading corrections at large $\Delta$.
We will argue that
\be\label{result}
F_\Delta^n =\left(1+ {\Delta_\lambda(\Delta_\lambda-1)\over 2\Delta} n+{\cal O}(\Delta^{-2})\right)p(n)
\ee
As a warmup, let us first reconsider our above example in more detail.
We have
\bea\
\langle \Delta | L_m \phi_\lambda(z) L_{-m}|\Delta\rangle
&=& 2m \Delta \langle \Delta |\phi(z)|\Delta\rangle
+ \langle \Delta | \left[L_m, [\phi(z), L_{-m}]\right]|\Delta\rangle
\eea
The first term dominates at large $\Delta$ and gives the Gram matrix described above.  The second term can be easily computed using
\be\label{Lphi}
[L_n ,\phi_\lambda(z)] = \Delta_\lambda(n+1)z^n \phi_\lambda(z)+z^{n+1}\p \phi_\lambda(z)
\ee
and
\be
\langle\Delta |\phi_\lambda(z)|\Delta\rangle = \langle\Delta |\phi_\lambda(1)|\Delta\rangle z^{-\Delta_\lambda}
\ee
to give
\be
\langle\Delta| [[L_m,[L_{-m}, \phi_\lambda(z)]|\Delta\rangle =
m^2 {{\Delta_\lambda}(\Delta_\lambda-1)}
\langle\Delta |\phi_\lambda(z)|\Delta\rangle
\ee
We see that the normalized three-point coefficient is
\be\label{eqz}
{\langle \Delta | L_m \phi_\lambda(z) L_{-m}|\Delta\rangle\over
	  \langle \Delta | L_m L_{-m}|\Delta\rangle}
= \left(1+ m {{\Delta_\lambda}(\Delta_\lambda-1)\over 2\Delta}  \right) \langle \Delta |\phi(z)|\Delta\rangle
\ee
We recognize in this formula the first subleading term in equation \rref{result}.

We will now argue that the above computation applies, at the desired order in $\Delta^{-1}$, to compute the contribution of a general descendant state. We wish to imagine a contribution of the form
\be
\langle \Delta |\left(\prod L_{i}^{n_i}\right) \phi(z)
\left(\prod L_{-j}^{m_j}\right)|\Delta\rangle [B^n_\Delta]^{MN}
\ee
As above, we compute this at large $\Delta$ by commuting the $L_i$ to the right.  The term with no commutators gives the leading contribution.   We can't have a single commutator, since there would be a mismatch in the levels of Virasoro generators on the two sides of $\phi_\lambda$, which would give zero.  So the first subleading contribution comes from a double commutator, which will be subleading by a factor of $1/\Delta$.
In fact, the leading contribution will come from the diagonal terms with $M=N$.  This is because after extracting the double commutator, unless $M=N$ there will be a mismatch in the types of Virasoro generator left on the two sides of $\phi_\lambda$; this would lead to fewer powers of $\Delta$ coming from $[L_n,L_{-n}]$ commutators.

Turning to the Gram matrix, we need the diagonal components of the inverse Gram matrix. These entries are just the inverse of the entries of the Gram matrix (to leading order in $\Delta$):
\be
    [B^n_\Delta]^{NN} = \left(\langle\Delta|\left(\prod L_{i}^{n_i}\right)
    \left(\prod L_{-j}^{n_j}\right) |\Delta\rangle\right)^{-1} \left(1+ {\cal O}(\Delta^{-1})\right)
    \ee That is, to leading order in $\Delta$ we can just set the off-diagonal terms in the Gram matrix to $0$. It's easy to verify this claim explicitly at level $2$.  This is because the leading contribution to the determinant of the full matrix and the minors just comes from the diagonal terms.

However, since in our computation we are only extracting a single double commutator, the computation proceeds exactly as in \rref{eqz}.  We have
\be
\langle \Delta |\left(\prod L_{-i}^{n_i}\right) \phi(z)
\left(\prod L_{j}^{n_j}\right)|\Delta\rangle [B^n_\Delta]^{NN}=
\left(1+ n {{\Delta_\lambda}(\Delta_\lambda-1)\over 2\Delta} +\dots  \right)
 \langle\Delta|\phi(z)|\Delta \rangle
\ee
which implies \rref{result}.

\subsection{Validity of large $\Delta$ asymptotics}

In the previous subsection we derived the following result for the large $\Delta$ expansion of the torus one-point block
\be\label{torexp}
F_\Delta(q)=\sum_{n=0}^\infty F_\Delta^n q^n = \sum_{n=0}^\infty\left(1+ {\Delta_\lambda(\Delta_\lambda-1)\over 2\Delta} n+{\cal O}(\Delta^{-2})\right)p(n)q^n
\ee
We now ask when it valid to keep just the leading term in the $1/\Delta$ expansion.   The issue is of course that no matter how large $\Delta$ is,  the sum over $n$ means that $n/\Delta$ grows arbitrarily large.  For the ${\cal O}(\Delta^{-1})$ term to be suppressed relative to the leading term we need for $q$ to be sufficiently small so that the sum is effectively cut off at $n \ll \Delta$.    To make this precise, we use the Hardy-Ramanujan formula
\be
 p(n) \sim e^{\pi \sqrt{2n\over 3}}~,\quad n \rightarrow \infty~.
\ee
Therefore $\sum_{n=0}^\infty p(n)q^n$ has a saddle point value of $n$ given by $n_* = \pi^2/6\beta^2$, where $q=e^{-\beta}$ as usual. Self-consistency requires $n_* \gg 1$ and  hence $\beta \ll 1$.   Under these conditions we have $n_*/\Delta \ll 1$ provided $\beta^2 \Delta \gg 1$.   This leads us to the desired result
\be
F_\Delta(q)  \approx \sum_{n=0}^\infty p(n)q^n = \prod_{n=1}^\infty {1\over 1-q^n}~,\quad {\rm for}\quad  \beta \ll 1 ~~{\rm and}~~  \beta^2\Delta \gg 1~.
\ee

\bibliographystyle{ssg}
\bibliography{biblio}

\begingroup\raggedright\begin{thebibliography}{10}

\bibitem{Belavin:1984vu} 
A.~A.~Belavin, A.~M.~Polyakov and A.~B.~Zamolodchikov,
``Infinite Conformal Symmetry in Two-Dimensional Quantum Field Theory,''
Nucl.\ Phys.\ B {\bf 241}, 333 (1984).

\bibitem{Cappelli:1986hf} 
A.~Cappelli, C.~Itzykson and J.~B.~Zuber,
``Modular Invariant Partition Functions in Two-Dimensions,''
Nucl.\ Phys.\ B {\bf 280}, 445 (1987).

\bibitem{Moore:1988uz} 
G.~W.~Moore and N.~Seiberg,
``Polynomial Equations for Rational Conformal Field Theories,''
Phys.\ Lett.\ B {\bf 212}, 451 (1988).

\bibitem{Sonoda:1988fq} 
H.~Sonoda,
``Sewing Conformal Field Theories. 2.,''
Nucl.\ Phys.\ B {\bf 311}, 417 (1988).

\bibitem{Rychkov:2016iqz}
S.~Rychkov, ``{EPFL Lectures on Conformal Field Theory in D$\geq $ 3
  Dimensions},'' \href{http://xxx.lanl.gov/abs/1601.05000}{{\tt 1601.05000}}.

\bibitem{Simmons-Duffin:2016gjk}
D.~Simmons-Duffin, ``{TASI Lectures on the Conformal Bootstrap},''
  \href{http://xxx.lanl.gov/abs/1602.07982}{{\tt 1602.07982}}.

\bibitem{Cardy:1986ie}
J.~L. Cardy, ``{Operator Content of Two-Dimensional Conformally Invariant
  Theories},'' {\em Nucl. Phys.} {\bf B270} (1986) 186--204.

\bibitem{Maldacena:1997re} 
J.~M.~Maldacena,
``The Large N limit of superconformal field theories and supergravity,''
Int.\ J.\ Theor.\ Phys.\  {\bf 38}, 1113 (1999),
\href{http://xxx.lanl.gov/abs/hep-th/9711200}{{\tt
hep-th/9711200}}.

\bibitem{Strominger:1997eq}
A.~Strominger, ``{Black hole entropy from near horizon microstates},'' {\em
  JHEP} {\bf 02} (1998) 009, \href{http://xxx.lanl.gov/abs/hep-th/9712251}{{\tt
  hep-th/9712251}}.

\bibitem{Hadasz:2009db}
L.~Hadasz, Z.~Jaskolski, and P.~Suchanek, ``{Recursive representation of the
  torus 1-point conformal block},'' {\em JHEP} {\bf 01} (2010) 063,
  \href{http://xxx.lanl.gov/abs/0911.2353}{{\tt 0911.2353}}.

\bibitem{Polchinski:1998rq}
J.~Polchinski, {\em {String theory. Vol. 1: An introduction to the bosonic
  string}}.
\newblock Cambridge University Press, 2007.

\bibitem{Carlip:2000nv} 
S.~Carlip,
``Logarithmic corrections to black hole entropy from the Cardy formula,''
Class.\ Quant.\ Grav.\  {\bf 17}, 4175 (2000),
\href{http://xxx.lanl.gov/abs/gr-qc/0005017}{{\tt
gr-qc/0005017}}.

\bibitem{Hellerman:2009bu} 
S.~Hellerman,
``A Universal Inequality for CFT and Quantum Gravity,''
JHEP {\bf 1108}, 130 (2011),
  \href{http://xxx.lanl.gov/abs/0902.2790}{{\tt 0902.2790}}.
  
%

\bibitem{Hartman:2014oaa}
T.~Hartman, C.~A. Keller, and B.~Stoica, ``{Universal Spectrum of 2d Conformal
  Field Theory in the Large c Limit},'' {\em JHEP} {\bf 09} (2014) 118,
  \href{http://xxx.lanl.gov/abs/1405.5137}{{\tt 1405.5137}}.

\bibitem{Friedan:2013cba} 
D.~Friedan and C.~A.~Keller,
``Constraints on 2d CFT partition functions,''
JHEP {\bf 1310}, 180 (2013),
  \href{http://xxx.lanl.gov/abs/1307.6562}{{\tt 1307.6562}}.

\bibitem{Banados:1992wn} 
M.~Banados, C.~Teitelboim and J.~Zanelli,
Phys.\ Rev.\ Lett.\  {\bf 69}, 1849 (1992),
\href{http://xxx.lanl.gov/abs/hep-th/9204099}{{\tt
hep-th/9204099}}.

\bibitem{Alkalaev:2016ptm} 
K.~B.~Alkalaev and V.~A.~Belavin,
JHEP {\bf 1606}, 183 (2016),
  \href{http://xxx.lanl.gov/abs/1603.08440}{{\tt 1603.08440}}.

\bibitem{Deser:1983tn}
S.~Deser, R.~Jackiw, and G.~'t~Hooft, ``{Three-Dimensional Einstein Gravity:
  Dynamics of Flat Space},'' {\em Annals Phys.} {\bf 152} (1984) 220.

\bibitem{Deser:1983nh}
S.~Deser and R.~Jackiw, ``{Three-Dimensional Cosmological Gravity: Dynamics of
  Constant Curvature},'' {\em Annals Phys.} {\bf 153} (1984) 405--416.

\bibitem{D'Hoker:2002aw}
E.~D'Hoker and D.~Z. Freedman, ``{Supersymmetric gauge theories and the AdS /
  CFT correspondence},'' in {\em {Strings, Branes and Extra Dimensions: TASI
  2001: Proceedings}}, pp.~3--158, 2002,
\newblock \href{http://xxx.lanl.gov/abs/hep-th/0201253}{{\tt hep-th/0201253}}.

\bibitem{KeskiVakkuri:1998nw}
E.~Keski-Vakkuri, ``{Bulk and boundary dynamics in BTZ black holes},'' {\em
  Phys. Rev.} {\bf D59} (1999) 104001,
  \href{http://xxx.lanl.gov/abs/hep-th/9808037}{{\tt hep-th/9808037}}.

\bibitem{Zamo:lectures}
A.~Zamolodchikov, ``{ Conformal symmetry in two-dimensional space: recursion
  representation of conformal block},'' in {\em { Theoretical and Mathematical
  Physics, Vol. 73, Issue 1, pp 1088-1093}}, 1987.

\bibitem{zamo}
A.~Zamolodchikov, ``{Conformal symmetry in two-dimensions: An explicit
  recurrence formula for the conformal partial wave amplitude},'' {\em
  Commun.Math.Phys.} {\bf 96} (1984) 419--422.


\bibitem{Maloney:2007ud} 
A.~Maloney and E.~Witten,
JHEP {\bf 1002}, 029 (2010),
\href{http://xxx.lanl.gov/abs/0712.0155}{{\tt 0712.0155}}.

\bibitem{Keller:2014xba} 
C.~A.~Keller and A.~Maloney,
``Poincare Series, 3D Gravity and CFT Spectroscopy,''
JHEP {\bf 1502}, 080 (2015),
\href{http://xxx.lanl.gov/abs/1407.6008}{{\tt 1407.6008}}.

\bibitem{Benjamin:2016aww} 
N.~Benjamin, E.~Dyer, A.~L.~Fitzpatrick, A.~Maloney and E.~Perlmutter,
``Small Black Holes and Near-Extremal CFTs,''
JHEP {\bf 1608}, 023 (2016),
\href{http://xxx.lanl.gov/abs/1603.08524}{{\tt 1603.08524}}.


\end{thebibliography}\endgroup

\end{document}